\documentclass[prb,twocolumn,superscriptaddress,floatfix]{revtex4}
\usepackage[dvips]{graphicx}
\begin{document}
\title{Electric field driven donor-based charge qubits in semiconductors} 
\author{Belita Koiller}
\affiliation{Condensed Matter Theory Center, Department of Physics,
University of Maryland, College Park, MD 20742-4111}
\affiliation{Instituto de F{\'\i}sica, Universidade Federal do Rio de
Janeiro, 21945, Rio de Janeiro, Brazil}
\author{Xuedong Hu}
\affiliation{Department of Physics, University at Buffalo, the State
University of New York, Buffalo, NY 14260-1500}
\author{S. Das Sarma}
\affiliation{Condensed Matter Theory Center, Department of Physics,
University of Maryland, College Park, MD 20742-4111}
\date{\today}

\begin{abstract}
We investigate theoretically donor-based charge qubit operation driven by external
electric fields. The basic physics of the problem is presented by considering a single electron bound to a shallow-donor pair in GaAs: This system is closely related to the homopolar molecular ion H$_2^+$. In the case of Si, heteropolar configurations such as PSb$^+$ pairs are also considered.  For both homopolar and heteropolar pairs, the
multivalley conduction band structure of Si leads to short-period
oscillations of the tunnel-coupling strength as a function of the inter-donor
relative position. However, for any fixed donor configuration, the response
of the bound electron to a uniform electric field in Si is  qualitatively
very similar to the GaAs case, with no valley quantum interference-related
effects, leading to the conclusion that electric field driven coherent manipulation of donor-based charge qubits is feasible in semiconductors.
\end{abstract}
\maketitle


Scalability constitutes one of the most attractive aspects of solid state
quantum computer (QC) proposals.  Group-IV or III-V semiconductor
nanostructure-based QC architectures further benefit from the existence of a
vast semiconductor microelectronics infrastructure.  Semiconductor QC
architectures typically rely on the electronic bound states of artificial
atoms or molecules embedded in semiconductors, where the electron confinement
potential is provided by quantum dots (self-assembled or gate-defined) or
shallow donor impurities.  Information encoded in the electron spin or
orbital state is processed through electronic charge manipulation by
externally applied electric
fields.\cite{Exch,Kane,Vrijen,Tanamoto99,Tanamoto00,fedichkin00,skinner03,hollenberg1} 
Single electron coherent effects in GaAs have been demonstrated for electrons
bound to donor impurities \cite{Cole00} as well as electrons in double
quantum dots.\cite{Hayashi,petta04} More recently, coherent time evolution of
electrons in an isolated double quantum dot in Si was also
reported.\cite{gorman05} 

Donor-based charge qubits are defined in close analogy with
double dot charge qubits,\cite{Hayashi} consisting of a singly ionized donor
pair $AB^+$.\cite{hollenberg1} The electron confining potential, usually
taken to be parabolic in simple quantum dot models, becomes hydrogenic
for donors.  
We have analyzed the tunnel coupling of P$_2^+$ charge qubits in
Si in Ref.~\onlinecite{hucharge}, where it was shown that the multivalley
structure of the Si conduction band leads to an oscillatory and anisotropic 
behavior of the tunnel-coupling strength as a function of the donor pair
relative position, favoring near-gap-closing cases.  Nonetheless, charge
qubit manipulation by electric fields performed in a Si double dot
\cite{gorman05} shows no indication of band-interference related phenomena in
the tunnel coupling when compared to similar experiments performed in GaAs
double dots.\cite{Hayashi}  It is thus important to clarify whether the
experimental observation indicates that in quantum dots the valley
interference effect is absent, or that the electronic response to the
external electric field is not sensitive to the valley interference, like the
hyperfine coupling strength in Si:P system.\cite{martins04}  

In this Communication we investigate the electric field manipulation 
of donor-based charge qubits in Si. For a better physical insight, it is instructive to consider initially donor-pair molecular ions in GaAs. Although shallow donor pairs in GaAs have simple and easy-to-control properties, sample preparation in GaAs poses an obvious difficulty that substitutional impurities at Ga or As sites will have entirely different behaviors (e.g. group-IV elements act as donors when replacing Ga, but acceptors when replacing As).  
Therefore the GaAs results presented here should be taken as a benchmark over which Si band effects are more clearly assessed.  Furthermore, shallow donor binding energies in Si present a much wider distribution of values compared to GaAs, ranging from $43$ meV for Sb to $71$ meV for Bi.  Thus the optimal coupling of different donor species in Si (forming an heteropolar $AB^+$
molecular ion, which is feasible experimentally\cite{schenkel05}) would require the assistance of an external electric field, which again poses the interesting question on whether valley interference in these double donor systems might lead to any difficulty in the charge manipulation by the external field. Our goal here is to assess the prospect for the coherent manipulation of donor-based Si charge qubits, similar to that achieved in GaAs\cite{Hayashi,petta04} and Si \cite{gorman05} quantum dot based systems, in light of the known\cite{hucharge} problem associated with the quantum interference among the valleys in the Si donor-based systems.

We model shallow donors in semiconductors within the effective mass
approximation (EMA),\cite{Pantelides} where the electronic properties of the
host material are described by a few parameters, usually taken to be the
static dielectric constant $\varepsilon$ and the effective masses at the
conduction-band edge. The case of GaAs is particularly simple theoretically,
since the conduction band minimum at ${\bf k}=0$ is non-degenerate and
isotropic, characterized by an effective mass $m^*=0.067 m_0$, where $m_0$ is
the free electron mass, and $\varepsilon = 12.56$. We consider group-IV (VI)
impurities replacing the group-III (V) element Ga (As), leading to a donor
state, i.e., an additional electron in the system bound to a hydrogenic
potential $U_i({\bf r}) = -e^2/(\varepsilon |{\bf r}-{\bf R}_i|)$ centered at
the impurity site ${\bf R}_i$.  The one-electron Hamiltonian is written as $H
= H_{\rm GaAs} + U_i({\bf r})$, where the first term describes the bulk
semiconductor material (kinetic energy and periodic potential).  Within EMA
the electron bound to the donor is
assumed to be highly delocalized in real space (with respect to the lattice
parameter), thus strongly localized in reciprocal space.  Additional
approximations\cite{Pantelides} allow the electron wave function to be
written as $\psi_i = F_i({\bf r})\phi_{\bf k}({\bf r})$, the product of a
slowly varying envelope function $F$ by the band-edge Bloch state $\phi_{\bf
k}$ which, in the case of GaAs, reduces to the rapidly varying periodic part
of the Bloch function $u_0$. The envelope function is the solution of the EMA
Hamiltonian $H_i=-\hbar^2 \bigtriangledown^2/(2m^*) + U_i({\bf r})$, leading
to $F_i({\bf r}) = [1/\sqrt{\pi(a^*)^3}] \exp(-|{\bf r}-{\bf R}_i|/a^*)$ with
effective Bohr radius $a^*=\varepsilon(m_0/m^*)a_0\approx 10$ nm and the
eigen-energy for the donor state with respect to the bottom of the conduction
band $E_D = -[m^*/(\varepsilon^2 m_0)]{\rm Ry} \approx -6$ meV, in excellent
agreement with shallow donor binding energies in GaAs 
(see Table\ref{table}).

\begin{figure}
\begin{center}
\includegraphics[width=85mm]
{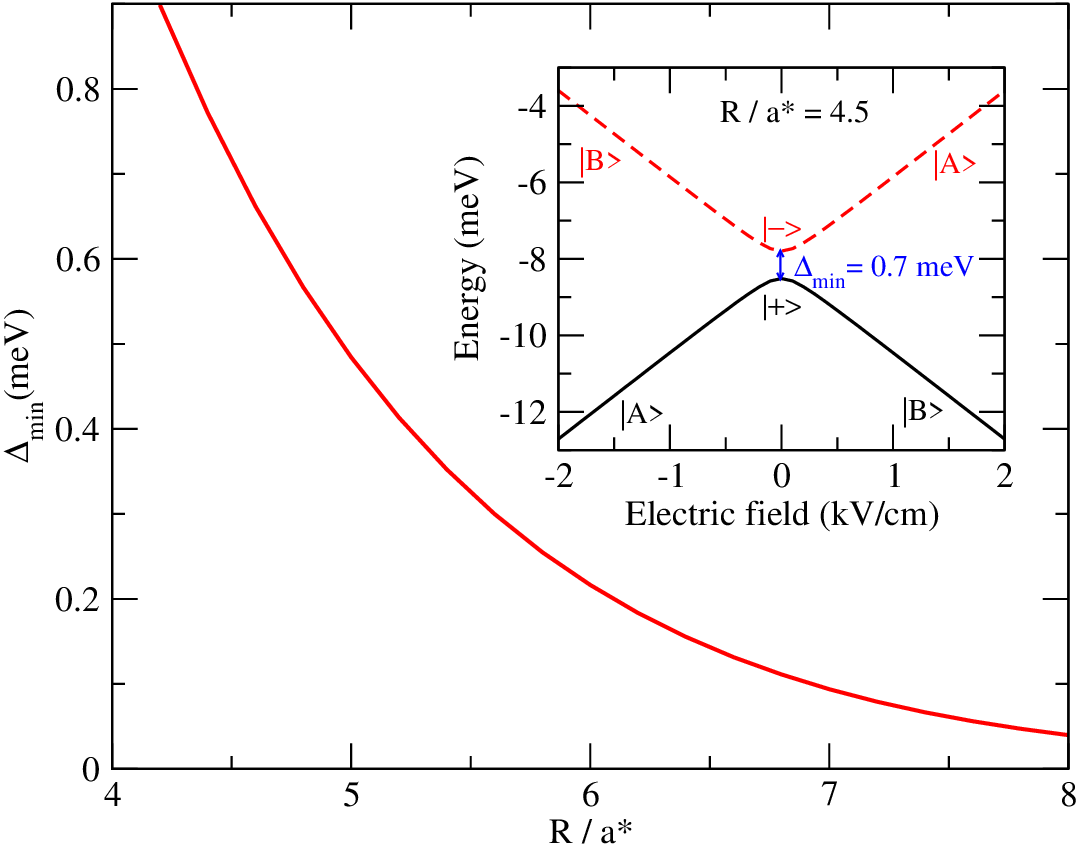}
\end{center}
\protect\caption[anti crossing for GaAs]
{\sloppy{
(Color online) Electric-field driven level anticrossing for a single electron
bound to a shallow donor pair in GaAs. The inset shows the calculated
eigen-energies (the solid and dashed lines give the ground and first excited
state respectively) versus applied axial field for interdonor distance
$R=4.5\, a^*$. As the field increases, the ground state changes from being
spatially localized around donor A to B. Level anticrossing occurs at
${\cal E}_c=0$: In this situation the eigenstates are the symmetric and
anti-symmetric superpositions $|+\rangle$ and $|-\rangle$ (see text). The
main graph gives the gap at anticrossing, $\Delta_{\rm min}$, versus $R$.   
}}
\label{fig:gaas}
\end{figure}

\begin{table}
\caption{\label{table}Shallow donor parameters for GaAs and Si: $E_D$ is the
ground state eigen-energy with respect to the conduction band edge, and $a^*$
is the effective Bohr radius.}
\begin{ruledtabular}
\begin{tabular}{c|c|cccc}
&GaAs \footnote[1]{Effective mass approximation.}& Si \footnote[2]{From
Ref.~\onlinecite{ningsah71}, where $E_D$ is fitted to experiment and $a^*$ is
calculated within the multivalley spherical band approximation.} &&& \\
& & Sb & P & As & Bi \\ \hline
$E_D$ (meV) &
$-5.8$ \footnote[3]{Experimental values of $-E_D$ for C, Si, Ge, S and Se
donors
in GaAs are, respectively, 5.9, 5.8, 5.9, 6.9 and 5.8 meV (from
Ref.~\onlinecite{madelung}).} 
& $-43$ & $-45$ & $-54$ & $-71$ \\
$a^*$(nm) & 9.9 & 1.32 & 1.22 & 0.97 & 0.74\\
\end{tabular}
\end{ruledtabular}
\end{table}

The Hamiltonian for a single electron bound to an ${A-B}$ donor pair at sites
${\bf R}_A$ and ${\bf R}_B$ in GaAs is 
\begin{equation}
H_{A-B}({\cal E}) = H_{\rm GaAs} + U_A + U_B + |e| \mathcal{E} \, {\bf \hat
s}\cdot
{\bf r} \,,
\label{eq:ha}
\end{equation}
where the last term accounts for an external axial field ${\bf {E}}={\cal
E}{\bf\, \hat
s }$, with ${\bf \hat s} = {\bf R}/R$ and ${\bf R} = {\bf R}_B -{\bf R}_A$.
For large enough $R$ $(R \gtrsim 4 a^*)$ this problem may be solved in
analogy with the LCAO solution for the H$_2^+$ molecular ion.\cite{slater}
The Hamiltonian is written in the non-orthogonal basis set
$\{\psi_A,\psi_B\}$, which amounts to writing an EMA Hamiltonian in the
$\{F_A,F_B\}$ basis set, since the periodic part of the Bloch function may be
ignored in practice.\cite{wellard03,KCHD} All matrix elements are integrated
analytically by separation of variables using spheroidal coordinates:\cite{notehetero}
$\lambda=(r_A+r_B)/R$ and $\mu=(r_A-r_B)/R$, with $r_i=|{\bf r} - {\bf
R_i}|$, $i=A,~B$. The electric field related term, not considered in
Ref.~\onlinecite{slater}, is transformed here through the relation ${\bf \hat
s}\cdot {\bf r} = \lambda \mu R/2$. 

Figure~\ref{fig:gaas} summarizes our results for donor charge qubits in GaAs.
The inset gives the characteristic level anticrossing of the qubit states,
driven here by the axial electric field, with the minimum gap $\Delta_{\rm
min}$ occuring at the anticrossing field ${\cal E}_c=0$. Far from the
anticrossing
region, the eigenstates correspond to the electron localized around one of
the individual donors, which we represent by $|A\rangle$ and $|B\rangle$
states, while for ${\cal E} = {\cal E}_c$ the ground state is the symmetric
superposition
$|+\rangle = (|A\rangle+|B\rangle)/\sqrt{2}$ and the first excited state is
the anti-symmetric superposition $|-\rangle =
(|A\rangle-|B\rangle)/\sqrt{2}$. As the interdonor distance $R$ increases,
$\Delta_{\rm min}$ decreases smoothly, ranging typically from one to two
orders of magnitude smaller than the single donor binding energy as $R$
varies  from 4$a^*$ to 8$a^*$.

\begin{figure}
\includegraphics[width=78mm]
{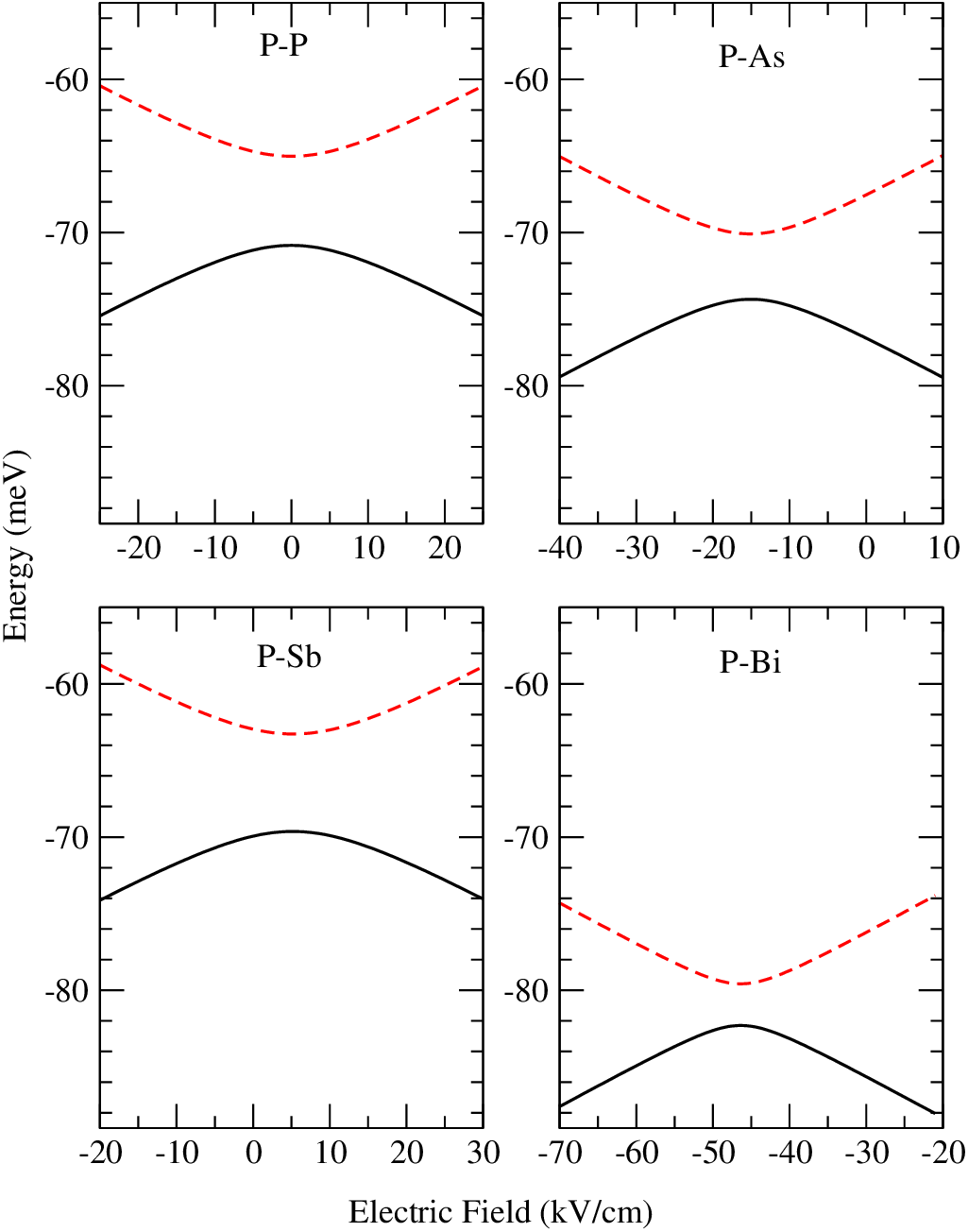}
\protect\caption[4 frames - level anticrossing]
{\sloppy{(Color online) Electric-field driven level anticrossing for the
indicated $A-B$ donor pairs forming an $AB^+$ molecular ion. The donors are
positioned a distance $R=4.5 \, a^*_{\rm P}$ apart, along the [110] Si
lattice direction, and the field is also applied along this direction. For
the heteropolar cases, a positive field moves the electron away from the P
site.}}
\label{fig:anticross}
\end{figure}

The theoretical description of shallow donors in Si is more complex than in
GaAs due to the six-fold degeneracy of the conduction band minimum of Si,
corresponding to non-zero values of ${\bf k}_\mu$ with ${\mu=1,\ldots 6}$.
Minima are located in reciprocal space 0.85 of the way along the $\Gamma$-X
line.  The energy dispersion near each of the band minimum is anisotropic,
leading to different effective masses for the longitudinal and transverse
directions along each of the $\Gamma$-X lines.  The tunnel coupling of P$_2^+$
homopolar charge qubits in Si was investigated in detail in
Ref.~\onlinecite{hucharge}, with the single donor
wavefunctions taken in the Kohn-Luttinger (KL) EMA form,\cite{Pantelides}
i.e., for a donor at ${\bf R}_A$,
\begin{equation}
\psi_A ({\bf r}) = {1\over \sqrt{6}} \sum_{\mu = 1}^6   F_{\mu}({\bf r}-{\bf
R}_A) u_\mu({\bf r}) e^{i {\bf k}_{\mu}\cdot ({\bf r}-{\bf R}_A)}~,
\label{eq:sim}
\end{equation}
where the envelope functions $F_\mu({\bf r})$ are deformed $1S$ hydrogenic
orbitals. With respect to the GaAs results in Fig.~\ref{fig:gaas}, the most
striking difference is the oscillatory and anisotropic behavior of
$\Delta_{min}$ vs $\bf R$ for P$_2^+$ in Si. It is well
established\cite{KCHD,wellard03,hucharge} that this behavior arises from
interference between the plane-wave parts of the Bloch functions, which are
pinned to the respective donor sites. 

For the heteropolar donor pairs, the multivalley character of the donor
electron wavefunction is accounted for in the form of Eq.~(\ref{eq:sim}).  We
adopt an additional simplification assuming isotropic envelope functions:
$F_\mu ({\bf r}) = [1/\sqrt{\pi(a_D^*)^3}] \exp(-|{\bf r}-{\bf R}_i|/a_D^*)$,
where $D$ labels the donor species. We also assume that, as for the case of
GaAs, the experimental donor binding energy satisfies $E_D = \langle \psi_D
|H_D|\psi_D \rangle$, where $H_D$ is the single-donor multivalley
Hamiltonian.\cite{hucharge} Our treatment of the donor pair under an electric
field follows closely the procedure outlined above for GaAs: The donor pair
Hamiltonian in the presence of an axial electric field, $H_{A-B}({\cal E})$,
is
written in the $\{\psi_A,\psi_B\}$ representation, in which one and
two-center integrals are readily calculated in spheroidal coordinates. 
We adopt here the values of the effective Bohr radii calculated by Ning and
Sah\cite{ningsah71} for the $1S(A1)$ state of the different donors within the
multivalley effective mass spherical-band approximation. This approach
involves a parametrization of the donor potential to fit the experimental
values of $E_D$. In Table \ref{table} we present the values of $a^*_D$ and
$E_D$ adopted here. Comparison with the KL effective Bohr radii (2.5 nm and
1.4 nm) shows that $a^*_D$ is probably underestimated. We expect our model to
be more reliable regarding the energy estimates, since all donor binding
energies are taken from experiment. 

\begin{figure}
\includegraphics[width=85mm]
{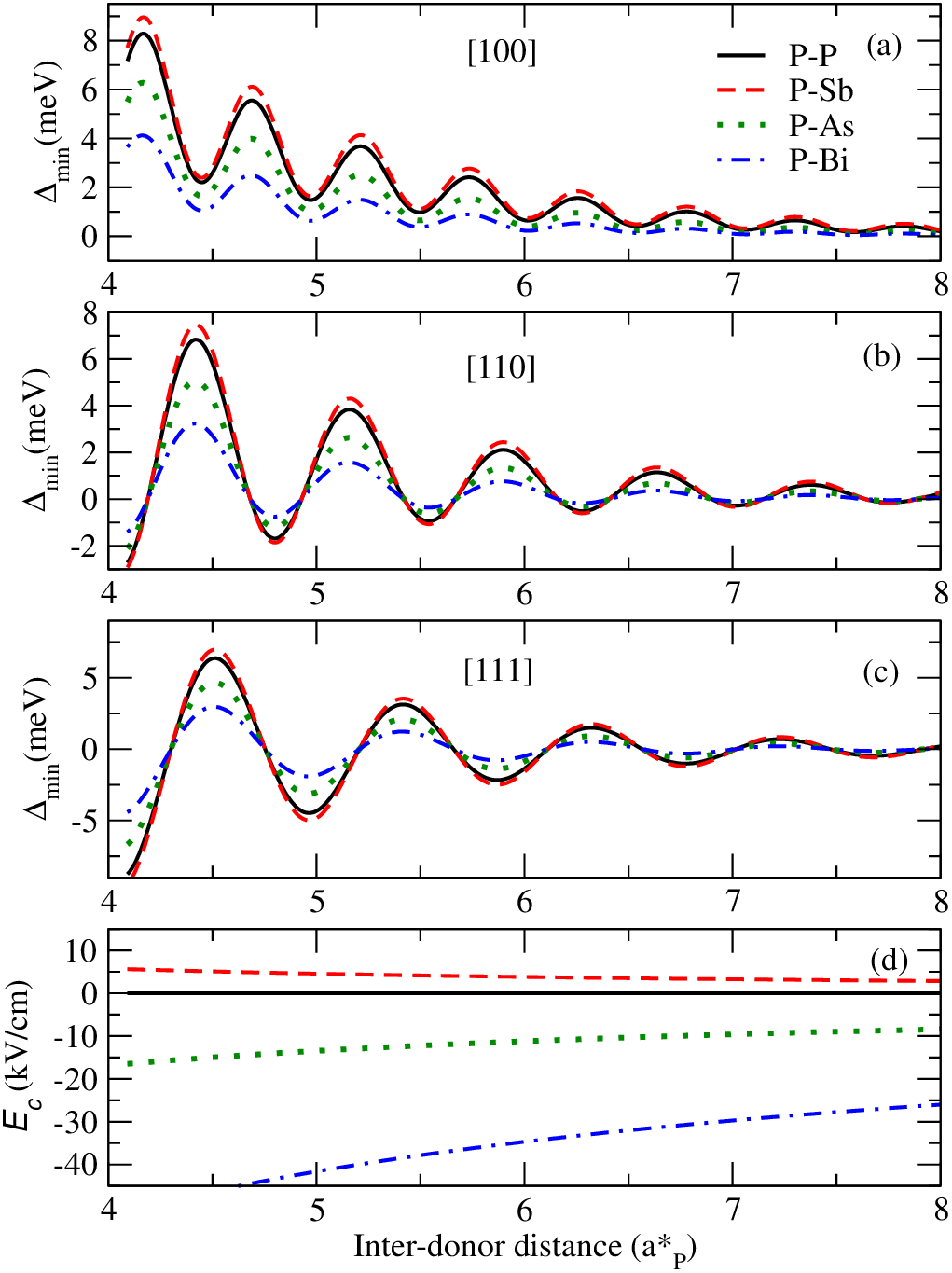}
\protect\caption[oscillatory Delta vs R and field]
{\sloppy{
(Color online) Calculated gap at anticrossing for biased $AB^+$ donor
molecular ions in Si versus interdonor relative position along the [100] (a),
[110] (b) and [111] (c) lattice direction.  Donor $A$ is P and donor $B$ is P
(solid lines), Sb (dashed lines), As (dotted lines) or Bi (dot-dash lines).
The axial electric field at anticrossing ${\cal E}_c$, given in frame (d), is
isotropic and depends on the donor species in the pair. Distances are given
in units of the P donor effective Bohr radius.
}}
\label{fig:fieldsi}
\end{figure}

Typical electric-field driven level anticrossing diagrams for donor molecular
ions $AB^+$  are given in Fig.~\ref{fig:anticross}, where one of the donors
$(A)$ is always chosen to be P. As compared to the P$_2^+$ case, the
eigen-energies for PSb$^+$ are higher while for PAs$^+$ and PBi$^+$ are
lower, reflecting respectively the shallower or deeper nature of donor
species $B$ with respect to P. Since shallower donors have larger
effective Bohr radii, yielding larger wavefunction
overlap, the gap at anti-crossing $\Delta_{min}$ also results to be larger
for shallower donors, following the eigen-energies trend. We note that, as
compared to
the GaAs case given in the inset of Fig.~\ref{fig:gaas}, no important
qualitative difference emerges for the electric field dependence of the
anti-crossing gap.  Of course, for the heteropolar cases,
level anticrossing occurs at a non-zero bias field ${\cal E}_c$. The
energy scales for GaAs are typically one order of magnitude smaller than for
Si, reflecting the difference in donor binding energies among these
materials.

Figure~\ref{fig:fieldsi} gives the gap at anticrossing versus interdonor
separation along three high-symmetry crystal directions in Si, as well as the
corresponding bias field ${\cal E}_c$. A major  difference between these
results and
the GaAs case is the oscillatory and anisotropic behavior of $\Delta_{min}$
vs $R$. Results for the P$_2^+$ molecular ion [solid lines in frames (a), (b)
and (c)] are in excellent qualitative agreement with those reported in
Ref.~\onlinecite{hucharge} (see Fig.~1 there). It is interesting to note
that, even considering isotropic envelope functions, anisotropic behavior is
obtained due to interference among the plane-wave parts (pinned at sites
${\bf R}_A$ and ${\bf R}_B$) of the six  Bloch functions, corresponding to
the wave vectors ${\bf k}_\mu$ at the Si conduction-band minima. The results
for heteropolar donor pairs, given by the dashed and dotted lines, follow the
same overall behavior as the P$_2^+$ case, showing that the oscillatory
tunnel-coupling strength is inherent to the Si host, and persists in
asymmetric molecular configurations. The same conclusions applicable to the
P$_2^+$ case,\cite{hucharge} in particular the tendency towards gap closing,
remain valid here. Comparison of the solid line with the dashed and dotted
lines in Fig.~\ref{fig:fieldsi}(a), (b) and (c) show that the coupling
strength increases (decreases) when shallower (deeper) donors form the pair,
as already noted in Fig.~\ref{fig:anticross}. The bias field at anticrossing
is plotted in Fig.~\ref{fig:fieldsi}(d), where positive bias is arbitrarily
chosen to lower the electronic potential at ${\bf R}_B$. 
Note that, for all heteropolar cases, $|{\cal E}_c|$ decreases monotonically 
as $R$ increases, which may be understood from the resonance condition 
$\langle \psi_A| H_{A-B}({\cal E}_c)|\psi_A\rangle = 
\langle \psi_B| H_{A-B}({\cal E}_c) |\psi_B\rangle$: Each matrix
element is taken between donor wavefunctions {\it pinned to a single site},
thus no quantum interference effects play a role, leading to the isotropic
non-oscillatory behavior of ${\cal E}_c(R)$ in Fig.~\ref{fig:fieldsi}(d). 
The resonance condition in the lowest order approximation leads 
to $|{\cal E}_c| = |E_A - E_B|/R$, which fits the data presented in
Fig.~\ref{fig:fieldsi}(d) quite well.

In this study we have shown that heteropolar donor pairs, like homopolar
pairs, suffer from sensitive dependence of antocrossing gap on interdonor
distance and orientation, resulting in a general narrowing of this gap, due
to the valley interference in the Si conduction band.  In the mean time,
as summarized in Figs.~\ref{fig:anticross} and \ref{fig:fieldsi}(d), we have
also shown that single electron response in a double donor system to an
external electric field is not affected by the valley interference in the Si
conduction band.  In other words, applying an external field will not cause
the tunnel coupling to exhibit oscillatory behavior as it does when
inter-donor distance is varied.  This conclusion is also applicable to Si
double dot charge qubit, no matter whether its state structure is similar to
that of a double donor in Si or double dot in GaAs.  Given a fixed
double-donor or double-dot geometry that has a large enough tunnel coupling,
coherent charge manipulation in Si with electric field should be quite
similar to that in GaAs, as recently demonstrated in
Ref.~\onlinecite{gorman05} via electrostatic pulse sequences in a Si double
quantum dot. We conclude that external electric field induced coherent manipulation of donor-based charge qubits should be experimentally possible {\it in spite of} the inter-valley quantum interference problem\cite{hucharge,KCHD} in silicon.

This work is supported by NSA, LPS, and ARO at the University of Maryland and
the University at Buffalo, and by Brazilian agencies CNPq, FUJB, FAPERJ,
PRONEX-MCT, and Instituto do Mil\^enio-CNPq.

\bibliography{charge}

\begin{thebibliography}{22}
\expandafter\ifx\csname natexlab\endcsname\relax\def\natexlab#1{#1}\fi
\expandafter\ifx\csname bibnamefont\endcsname\relax
  \def\bibnamefont#1{#1}\fi
\expandafter\ifx\csname bibfnamefont\endcsname\relax
  \def\bibfnamefont#1{#1}\fi
\expandafter\ifx\csname citenamefont\endcsname\relax
  \def\citenamefont#1{#1}\fi
\expandafter\ifx\csname url\endcsname\relax
  \def\url#1{\texttt{#1}}\fi
\expandafter\ifx\csname urlprefix\endcsname\relax\def\urlprefix{URL }\fi
\providecommand{\bibinfo}[2]{#2}
\providecommand{\eprint}[2][]{\url{#2}}

\bibitem[{\citenamefont{Loss and DiVincenzo}(1998)}]{Exch}
\bibinfo{author}{\bibfnamefont{D.}~\bibnamefont{Loss}} \bibnamefont{and}
  \bibinfo{author}{\bibfnamefont{D.P.} \bibnamefont{DiVincenzo}},
  \bibinfo{journal}{Phys. Rev. A} \textbf{\bibinfo{volume}{57}},
  \bibinfo{pages}{120} (\bibinfo{year}{1998}).

\bibitem[{\citenamefont{Kane}(1998)}]{Kane}
\bibinfo{author}{\bibfnamefont{B.~E.} \bibnamefont{Kane}},
  \bibinfo{journal}{Nature} \textbf{\bibinfo{volume}{393}},
  \bibinfo{pages}{133} (\bibinfo{year}{1998}).

\bibitem[{\citenamefont{Vrijen et~al.}(2000)\citenamefont{Vrijen, Yablonovitch,
  Wang, Jiang, Balandin, Roychowdhury, Mor, and DiVincenzo}}]{Vrijen}
\bibinfo{author}{\bibfnamefont{R.}~\bibnamefont{Vrijen}},
  \bibinfo{author}{\bibfnamefont{E.}~\bibnamefont{Yablonovitch}},
  \bibinfo{author}{\bibfnamefont{K.}~\bibnamefont{Wang}},
  \bibinfo{author}{\bibfnamefont{H.~W.} \bibnamefont{Jiang}},
  \bibinfo{author}{\bibfnamefont{A.}~\bibnamefont{Balandin}},
  \bibinfo{author}{\bibfnamefont{V.}~\bibnamefont{Roychowdhury}},
  \bibinfo{author}{\bibfnamefont{T.}~\bibnamefont{Mor}}, \bibnamefont{and}
  \bibinfo{author}{\bibfnamefont{D.P.}~\bibnamefont{DiVincenzo}},
  \bibinfo{journal}{Phys. Rev. A} \textbf{\bibinfo{volume}{62}},
  \bibinfo{pages}{012306} (\bibinfo{year}{2000}).

\bibitem[{\citenamefont{Tanamoto}(1999)}]{Tanamoto99}
\bibinfo{author}{\bibfnamefont{T.}~\bibnamefont{Tanamoto}},
  \bibinfo{journal}{Physica B} \textbf{\bibinfo{volume}{272}},
  \bibinfo{pages}{45} (\bibinfo{year}{1999}).

\bibitem[{\citenamefont{Tanamoto}(2000)}]{Tanamoto00}
\bibinfo{author}{\bibfnamefont{T.}~\bibnamefont{Tanamoto}},
  \bibinfo{journal}{Phys. Rev. A} \textbf{\bibinfo{volume}{61}},
  \bibinfo{pages}{022305} (\bibinfo{year}{2000}).

\bibitem[{\citenamefont{Fedichkin et~al.}(2000)\citenamefont{Fedichkin,
  Yanchenko, and Valiev}}]{fedichkin00}
\bibinfo{author}{\bibfnamefont{L.~E.} \bibnamefont{Fedichkin}},
  \bibinfo{author}{\bibfnamefont{M.}~\bibnamefont{Yanchenko}},
  \bibnamefont{and} \bibinfo{author}{\bibfnamefont{K.~A.}
  \bibnamefont{Valiev}}, \bibinfo{journal}{Nanotechnology}
  \textbf{\bibinfo{volume}{11}}, \bibinfo{pages}{387} (\bibinfo{year}{2000}).

\bibitem[{\citenamefont{Skinner et~al.}(2003)\citenamefont{Skinner, Davenport,
  and Kane}}]{skinner03}
\bibinfo{author}{\bibfnamefont{A.~J.} \bibnamefont{Skinner}},
  \bibinfo{author}{\bibfnamefont{M.~E.} \bibnamefont{Davenport}},
  \bibnamefont{and} \bibinfo{author}{\bibfnamefont{B.~E.} \bibnamefont{Kane}},
  \bibinfo{journal}{Phys. Rev. Lett.} \textbf{\bibinfo{volume}{90}},
  \bibinfo{pages}{087901} (\bibinfo{year}{2003}).

\bibitem[{\citenamefont{Hollenberg et~al.}(2004)\citenamefont{Hollenberg,
  Dzurak, Wellard, Hamilton, Reilly, Milburn, and Clark}}]{hollenberg1}
\bibinfo{author}{\bibfnamefont{L.~C.~L.} \bibnamefont{Hollenberg}},
  \bibinfo{author}{\bibfnamefont{A.~S.} \bibnamefont{Dzurak}},
  \bibinfo{author}{\bibfnamefont{C.~J.} \bibnamefont{Wellard}},
  \bibinfo{author}{\bibfnamefont{A.~R.} \bibnamefont{Hamilton}},
  \bibinfo{author}{\bibfnamefont{D.~J.} \bibnamefont{Reilly}},
  \bibinfo{author}{\bibfnamefont{G.~J.} \bibnamefont{Milburn}},
  \bibnamefont{and} \bibinfo{author}{\bibfnamefont{R.~G.} \bibnamefont{Clark}},
  \bibinfo{journal}{Phys. Rev. B} \textbf{\bibinfo{volume}{69}},
  \bibinfo{pages}{113301} (\bibinfo{year}{2004}).

\bibitem[{\citenamefont{Cole et~al.}(2000)\citenamefont{Cole, Williams, King,
  Sherwin, and Stanley}}]{Cole00}
\bibinfo{author}{\bibfnamefont{B.~E.} \bibnamefont{Cole}},
  \bibinfo{author}{\bibfnamefont{J.~B.} \bibnamefont{Williams}},
  \bibinfo{author}{\bibfnamefont{B.~T.} \bibnamefont{King}},
  \bibinfo{author}{\bibfnamefont{M.~S.} \bibnamefont{Sherwin}},
  \bibnamefont{and} \bibinfo{author}{\bibfnamefont{C.~R.}
  \bibnamefont{Stanley}}, \bibinfo{journal}{Nature}
  \textbf{\bibinfo{volume}{410}}, \bibinfo{pages}{60} (\bibinfo{year}{2000}).

\bibitem[{\citenamefont{Hayashi et~al.}(2003)\citenamefont{Hayashi, Fujusawa,
  Cheong, Jeong, and Hirayama}}]{Hayashi}
\bibinfo{author}{\bibfnamefont{T.}~\bibnamefont{Hayashi}},
  \bibinfo{author}{\bibfnamefont{T.}~\bibnamefont{Fujusawa}},
  \bibinfo{author}{\bibfnamefont{H.~D.} \bibnamefont{Cheong}},
  \bibinfo{author}{\bibfnamefont{Y.~H.} \bibnamefont{Jeong}}, \bibnamefont{and}
  \bibinfo{author}{\bibfnamefont{Y.}~\bibnamefont{Hirayama}},
  \bibinfo{journal}{Phys. Rev. Lett.} \textbf{\bibinfo{volume}{91}},
  \bibinfo{pages}{226804} (\bibinfo{year}{2003}).

\bibitem[{\citenamefont{Petta et~al.}(2004)\citenamefont{Petta, Johnson,
  Marcus, Hanson, and Gossard}}]{petta04}
\bibinfo{author}{\bibfnamefont{J.~R.} \bibnamefont{Petta}},
  \bibinfo{author}{\bibfnamefont{A.~C.} \bibnamefont{Johnson}},
  \bibinfo{author}{\bibfnamefont{C.~M.} \bibnamefont{Marcus}},
  \bibinfo{author}{\bibfnamefont{M.~P.} \bibnamefont{Hanson}},
  \bibnamefont{and} \bibinfo{author}{\bibfnamefont{A.~C.}
  \bibnamefont{Gossard}}, \bibinfo{journal}{Phys. Rev. Lett,}
  \textbf{\bibinfo{volume}{93}}, \bibinfo{pages}{186802}
  (\bibinfo{year}{2004}).

\bibitem[{\citenamefont{Gorman et~al.}(2005)\citenamefont{Gorman, Hasko, and
  Williams}}]{gorman05}
\bibinfo{author}{\bibfnamefont{J.}~\bibnamefont{Gorman}},
  \bibinfo{author}{\bibfnamefont{D.~G.} \bibnamefont{Hasko}}, \bibnamefont{and}
  \bibinfo{author}{\bibfnamefont{D.~A.} \bibnamefont{Williams}},
  \bibinfo{journal}{Phys. Rev. Lett,} \textbf{\bibinfo{volume}{95}},
  \bibinfo{pages}{090502} (\bibinfo{year}{2005}).

\bibitem[{\citenamefont{Hu et~al.}(2005)\citenamefont{Hu, Koiller, and {Das
  Sarma}}}]{hucharge}
\bibinfo{author}{\bibfnamefont{X.}~\bibnamefont{Hu}},
  \bibinfo{author}{\bibfnamefont{B.}~\bibnamefont{Koiller}}, \bibnamefont{and}
  \bibinfo{author}{\bibfnamefont{S.}~\bibnamefont{{Das Sarma}}},
  \bibinfo{journal}{Phys. Rev. B} \textbf{\bibinfo{volume}{71}},
  \bibinfo{pages}{235332} (\bibinfo{year}{2005}).

\bibitem[{\citenamefont{Martins et~al.}(2004)\citenamefont{Martins, Capaz, and
  Koiller}}]{martins04}
\bibinfo{author}{\bibfnamefont{A.~S.} \bibnamefont{Martins}},
  \bibinfo{author}{\bibfnamefont{R.~B.} \bibnamefont{Capaz}}, \bibnamefont{and}
  \bibinfo{author}{\bibfnamefont{B.}~\bibnamefont{Koiller}},
  \bibinfo{journal}{Phys. Rev. B} \textbf{\bibinfo{volume}{69}},
  \bibinfo{pages}{085320} (\bibinfo{year}{2004}).

\bibitem[{\citenamefont{Schenkel et~al.}(2005)\citenamefont{Schenkel,
  Tyryshkin, de~Sousa, Whaley, Bokor, J.~A.~Liddle, Persaud, Shangkuan,
  Charakov, and Lyon}}]{schenkel05}
\bibinfo{author}{\bibfnamefont{T.}~\bibnamefont{Schenkel}},
  \bibinfo{author}{\bibfnamefont{A.~M.} \bibnamefont{Tyryshkin}},
  \bibinfo{author}{\bibfnamefont{R.}~\bibnamefont{de~Sousa}},
  \bibinfo{author}{\bibfnamefont{K.~B.} \bibnamefont{Whaley}},
  \bibinfo{author}{\bibfnamefont{J.}~\bibnamefont{Bokor}},
  \bibinfo{author}{\bibnamefont{J.~A.~Liddle}},
  \bibinfo{author}{\bibfnamefont{A.}~\bibnamefont{Persaud}},
  \bibinfo{author}{\bibfnamefont{J.}~\bibnamefont{Shangkuan}},
  \bibinfo{author}{\bibfnamefont{I.}~\bibnamefont{Charakov}}, \bibnamefont{and}
  \bibinfo{author}{\bibfnamefont{S.~A.} \bibnamefont{Lyon}},
  \bibinfo{journal}{cond-mat/0507318}  (\bibinfo{year}{2005}).

\bibitem[{\citenamefont{Pantelides}(1978)}]{Pantelides}
\bibinfo{author}{\bibfnamefont{S.~T.} \bibnamefont{Pantelides}},
  \bibinfo{journal}{Rev. Mod. Phys.} \textbf{\bibinfo{volume}{50}}
  (\bibinfo{year}{1978}).

\bibitem[{\citenamefont{Ning and Sah}(1971)}]{ningsah71}
\bibinfo{author}{\bibfnamefont{T.~H.} \bibnamefont{Ning}} \bibnamefont{and}
  \bibinfo{author}{\bibfnamefont{C.~T.} \bibnamefont{Sah}},
  \bibinfo{journal}{Phys. Rev. B} \textbf{\bibinfo{volume}{4}},
  \bibinfo{pages}{3468} (\bibinfo{year}{1971}).

\bibitem[{\citenamefont{Madelung}(1996)}]{madelung}
\bibinfo{author}{\bibfnamefont{O.}~\bibnamefont{Madelung}},
  \emph{\bibinfo{title}{Semiconductors-Basic Data}}
  (\bibinfo{publisher}{Springer, Berlin}, \bibinfo{year}{1996}).

\bibitem[{\citenamefont{Slater}(1963)}]{slater}
\bibinfo{author}{\bibfnamefont{J.~C.} \bibnamefont{Slater}},
  \emph{\bibinfo{title}{Quantum Theory of Molecules and Solids}},
  vol.~\bibinfo{volume}{1} (\bibinfo{publisher}{McGraw-Hill, New York},
  \bibinfo{year}{1963}).

\bibitem[{\citenamefont{Wellard et~al.}(2003)\citenamefont{Wellard, Hollenberg,
  Parisoli, Kettle, Goan, McIntosh, and Jamieson}}]{wellard03}
\bibinfo{author}{\bibfnamefont{C.~J.} \bibnamefont{Wellard}},
  \bibinfo{author}{\bibfnamefont{L.~C.~L.} \bibnamefont{Hollenberg}},
  \bibinfo{author}{\bibfnamefont{F.}~\bibnamefont{Parisoli}},
  \bibinfo{author}{\bibfnamefont{L.}~\bibnamefont{Kettle}},
  \bibinfo{author}{\bibfnamefont{H.-S.} \bibnamefont{Goan}},
  \bibinfo{author}{\bibfnamefont{J.~A.~L.} \bibnamefont{McIntosh}},
  \bibnamefont{and} \bibinfo{author}{\bibfnamefont{D.~N.}
  \bibnamefont{Jamieson}}, \bibinfo{journal}{Phys. Rev. B}
  \textbf{\bibinfo{volume}{68}}, \bibinfo{pages}{195209}
  (\bibinfo{year}{2003}).

\bibitem[{\citenamefont{Koiller et~al.}(2004)\citenamefont{Koiller, Capaz, Hu,
  and {Das Sarma}}}]{KCHD}
\bibinfo{author}{\bibfnamefont{B.}~\bibnamefont{Koiller}},
  \bibinfo{author}{\bibfnamefont{R.~B.} \bibnamefont{Capaz}},
  \bibinfo{author}{\bibfnamefont{X.}~\bibnamefont{Hu}}, \bibnamefont{and}
  \bibinfo{author}{\bibfnamefont{S.}~\bibnamefont{{Das Sarma}}},
  \bibinfo{journal}{Phys. Rev. B} \textbf{\bibinfo{volume}{70}},
  \bibinfo{pages}{115207} (\bibinfo{year}{2004}).

\bibitem[{not()}]{notehetero}
\bibinfo{note}{This separation of variables also allows solving the problem for
  a pair of different donors, which is of no practical merit here since the
  binding energy of shallow donors in GaAs is essentially independent of the
  donor element.}

\end{thebibliography}

\end{document}